\documentclass{article}
\usepackage{multicol}
\usepackage{fullpage}
\begin{document}

\title{Mechanics: non-classical, non-quantum
}

\author{Elliott Francesco Tammaro\\ \footnotesize{Bryn Mawr College, Bryn Mawr, PA
\emph{e-mail:} etammaro@brynmawr.edu}}

\maketitle \footnote{This article appears in \emph{Foundations of Physics}}

\begin{abstract}
A non-classical, non-quantum theory, or NCQ, is any fully consistent theory that differs fundamentally from both the corresponding classical and quantum theories, while exhibiting certain features common to both. Such theories are of interest for two primary reasons. Firstly, NCQs arise prominently in semi-classical approximation schemes. Their formal study may yield improved approximation techniques in the near-classical regime. More importantly for the purposes of this note, it may be possible for NCQs to reproduce quantum results over experimentally tested regimes while having a well defined classical limit, and hence are viable alternative theories. We illustrate an NCQ by considering an explicit class of NCQ mechanics. Here this class will be arrived at via a natural generalization of classical mechanics formulated in terms of a probability density functional.

\end{abstract}

\section{Introduction}
\label{intro}
Quantum field theories, whose basis is formed by the special theory of relativity and quantum mechanics (QM), are some of the most well tested theories in history. However, remaining questions about renormalization and the failure to produce a quantum gravity theory (apart from attempts such as string theory, loops, etc), suggest that alternatives to the standard quantization prescription and/or Lorentz symmetry violations be studied. The success of non-relativistic QM, and a general lack of alternatives to standard quantization have acted to shift research in the direction of Lorentz violations. Indeed, there is an immense amount of experimental \cite{A8}-\cite{refA4}, and theoretical work \cite{refA5}-\cite{refIV} whose sole aim is the observation and explanation of a broken Lorentz symmetry (\emph{this list is by no means complete}). There remain no confirmed examples of Lorentz violations, even with the increasingly stringent limits placed on the Lorentz violating parameters in the Standard-Model Extension (SME), Robertson, or Mansouri-Sexl test theories, see \cite{A8}-\cite{refA4}.  

If one accepts the validity of Lorentz symmetry as at least a very good approximate symmetry, then a natural area of non-string theoretic research is the search for alternatives to the quantization prescription. In particular, one may study classes of theories that mimic, or effectively reproduce, the predictions of a quantum theory over experimentally tested regimes, but whose formalism may differ significantly. More specifically, viable alternative theories must meet two essential criteria. They must reproduce experimental results within regimes over which the quantum theory has been tested, and they must have a well defined classical limit. The benefit of considering such theories is that they are not rigidly bound to the formalism of quantum theories, and consequently, one might speculate that they may bypass some of the difficulties that face general QFT, and in particular quantum gravity$-$ i.e. renormalization and infinite vacuum energy. This could be accomplished because, as will be observed, the class of theories introduced herein are fully defined through a probability density functional, which is chosen integrably finite. The result of which is, naturally, that expectation values and transition probabilities are themselves finite quantities. As an entry into such a study we will consider non-classical, non-quantum theories, herein referred to as NCQs. By definition, they differ from their classical and quantum counterparts, while exhibiting certain features common to both.

In this note, we will not introduce a possible theoretical alternative to quantum mechanics, which remains a daunting task. Instead, we will introduce an NCQ theory of mechanics, with a classical limit, in order to illustrate and motivate the definition of an NCQ. In section \ref{sec:1} we will consider a reformulation of classical mechanics (CM) in terms of a probability density functional (PDF). In section \ref{sec:2} we will illustrate the PDF reformulation of classical mechanics by applying it to the one dimensional classical harmonic oscillator. In section \ref{sec:3} we will make use of this reformulation of CM to generate NCQ mechanics, and therein we will undertake a comparison between NCQ mechanics and QM/CM.  

\section{Probability Density Functional Formulation of Classical Mechanics}
\label{sec:1}

Consider a classical point particle interacting with some fixed external potential. We assume the equations of motion have a general solution of the form $\mathbf{x}_s(\alpha_1, \alpha_2,...\alpha_n;t)$ where $\{ \alpha_1,\alpha_2,...\alpha_n \}$ is the set of integration constants. Now define the following probability density functional: \begin{eqnarray}\label{equation 1} P_c[\mathbf{x}(t)]&\equiv& F[\mathbf{x}(t)]\int d\alpha_1 d\alpha_2...d\alpha_n \delta[\mathbf{x}(t)-\mathbf{x}_s(\alpha_1, \alpha_2,...\alpha_n;t)]\\ &=& \label{equation 2}F[\mathbf{x}(t)] \int d\alpha_1 d\alpha_2...d\alpha_n\label{I} \prod_t
\delta(\mathbf{x}(t)-\mathbf{x}_s(\alpha_1, \alpha_2,...\alpha_n;t))\end{eqnarray} where $\delta[...]$ is a delta functional, and where $\prod_t$ symbol represents discretizing $t$, taking the product as written, and finally taking the continuum limit. For all $\mathbf{x}(t)$ \begin{eqnarray*}P_c[\mathbf{x}(t)]&\in& \mathbf{R} \\ P_c[\mathbf{x}(t)]&\geq& 0\end{eqnarray*} and $F[\mathbf{x}(t)]$, a normalizing functional, is chosen to normalize $P_c[\mathbf{x}(t)]$ in the usual manner \begin{equation}\int_{\mbox{all }\mathbf{x}(t)} \mathcal{D}\mathbf{x}(t)P_c[\mathbf{x}(t)]=1\end{equation} where $\int \mathcal{D}\mathbf{x}(t)$ represents a path integral over trajectories, $\mathbf{x}(t)$ \cite{refPeskin}. We interpret $P_c[\mathbf{x}(t)]$ as follows: 
$\int_\mathcal{C}\mathcal{D}\mathbf{x}(t)P_c[\mathbf{x}(t)]$ is the probability that during a particular experiment the particle will be observed following a trajectory $\mathbf{x}(t)\in \mathcal{C}$. From this definition it follows that one may define the \emph{expectation} \emph{value} of an observable $O=O[\mathbf{x}(t)]$ in the usual way: \begin{equation}\label{expvalue}\langle O \rangle_c\equiv \int \mathcal{D}\mathbf{x}(t) O[\mathbf{x}(t)] P_c[\mathbf{x}(t)]\end{equation}

The form of $P_c[\mathbf{x}(t)]$ in (\ref{equation 1}) has been chosen such that $P_c[\mathbf{x}(t)]\not=0$ implies $\mathbf{x}(t)$ is a classically allowed trajectory. A \emph{classical} point particle system is equipped with equations of motion, which have as a solution set $S$ a set of trajectories $\mathbf{x}_s(\alpha_1, \alpha_2,...\alpha_n;t)$. Any program that allows for the calculation of $S$ will be considered a reformulation of classical mechanics. Consequently, the aforementioned formulation reproduces classical mechanics since $S$ is assembled by identifying trajectories, $\mathbf{x}(t)$, that satisfy $P_c[\mathbf{x}(t)]\not=0$. As one may expect $P_c[\mathbf{x}(t)]$ is expressible in terms of the equations of motion in an obvious way, see \cite{refI} for details. As a final note, $F[\mathbf{x}(t)]$ is allowed to vary as a functional because of the possibility to prepare particles with certain fixed initial conditions so that the probability distribution varies on the set of classically allowed trajectories. For example, if one forces $F[\mathbf{x}(t)]$ to factor into $N\delta(\mathbf{x}(t_0)-\mathbf{x}_0)$, with $N$ normalizing, then the trajectories are ``pinned" to the point $(\mathbf{x}_0, t_0)$. One may state in a precise manner that the equations of motion and initial conditions are necessary and sufficient data for the specification of the PDF. 

We note in passing that this reformulation makes explicit use of a path integral. In this way, it is reminiscent of the path integral formulation of QM or CM \cite{refFeynman} \cite{refI}. However, in contrast with the path integral formulation, which introduces a generating functional or transition amplitude that is calculated by performing a path integral over $e^{\frac{i}{\hbar}S}$ (where $S$ is the classical action), in this work the probability distribution functional is taken as the fundamental object of study. 

\section{Classical Harmonic Oscillator PDF in One Dimension}
\label{sec:2}
The equation of motion for the 1D H.O. is $\frac{d^2}{dt^2}x(t)=-\frac{k}{m}x(t)$. The general solution is \begin{equation}x(t)=A\sin(\omega t)+B\cos(\omega t)\end{equation} where $\omega$, the angular frequency, is $\sqrt{\frac{k}{m}}$. Following \ref{equation 1}, we see that the PDF for as of yet unfixed $F[x(t)]$ is \begin{equation}\label{HOPDF}P_{HO}[x(t)]=F[x(t)]\int dA dB \phantom{i}\delta[x(t)-A\sin(\omega t)-B\cos(\omega t)]\end{equation} 
Now we simultaneously satisfy initial conditions and normalize by fixing $F[x(t)]$. This is readily accomplished. Suppose we wish to enforce the initial conditions that $x(0)=x_0$ and $\dot{x}(0)=v_0$. Then, F[x(t)] must be proportional to $\delta(x(0)-x_0)\delta(\dot{x}(0)-v_0)$. We have that (\ref{HOPDF}) $=$ \begin{eqnarray*}N\delta(x(0)-x_0)\delta(\dot{x}(0)-v_0)\int dA dB \phantom{i}\delta[x(t)-A\sin(\omega t)-B\cos(\omega t)]\mathnormal{.}\end{eqnarray*} Finally, we deduce $N$ from normalization: \begin{eqnarray*}\int \mathcal{D}x(t) P_{HO}[x(t)]&=&N\int \mathcal{D}x(t)\delta(x(0)-x_0)\delta(\dot{x}(0)-v_0)\int dA dB \phantom{i}\delta[x(t)-A\sin(\omega t)-B\cos(\omega t)]\\ &=&N\int dA dB \int \mathcal{D}x(t)\delta(x(0)-x_0)\delta(\dot{x}(0)-v_0)\delta[x(t)-A\sin(\omega t)-B\cos(\omega t)]\end{eqnarray*} which yields, upon performing the path integral with the delta functional integrand, \begin{eqnarray*}&N&\int dA dB\delta(B-x_0)\delta(\omega A-v_0)=\frac{N}{\omega}=1\end{eqnarray*} Therefore, \begin{eqnarray*}P_{HO}[x(t)]=\omega \delta(x(0)-x_0)\delta(\dot{x}(0)-v_0)\int dA dB \phantom{i}\delta[x(t)-A\sin(\omega t)-B\cos(\omega t)]\end{eqnarray*} This gives explicit insight into how the PDF depends upon the choice of initial conditions.  

To further demonstrate the usefulness of this reformulation, let us calculate the expectation value of the energy using the PDF. Using \ref{expvalue} and the energy function, $E=\frac{1}{2}m\dot{x}(t)^2 + \frac{1}{2}m\omega^2x(t)^2$, we see that \begin{eqnarray*}\langle E \rangle_{HO}\equiv \int \mathcal{D}x(t) E(x(t)) P_{HO}[x(t)]=\int \mathcal{D}x(t)(\frac{1}{2}m\dot{x}(t)^2 + \frac{1}{2}m\omega^2x(t)^2) P_{HO}[x(t)]\end{eqnarray*} The calculation proceeds in a straightforward manner- the integrals are trivial because of the presence of the delta functions/functional. The final result is \begin{eqnarray*}\langle E \rangle_{HO}=\frac{1}{2}mv_0^2 + \frac{1}{2}m\omega^2x_0^2\end{eqnarray*} exactly as expected given the initial conditions. 

\section{NCQ Mechanics}
\label{sec:3}

Observe that in (\ref{equation 2}) appears the following product of delta functions \begin{equation}\prod_t
\delta(\mathbf{x}(t)-\mathbf{x}_s(\alpha_1, \alpha_2,...\alpha_n;t)).\end{equation} The delta functions act to restrict the PDF so that it is nonzero only on classically allowed trajectories, and hence are absolutely necessary for a classical interpretation. However, only the requirements that the delta functions are positive semidefinite and integrably finite are necessary for constructing a valid PDF. We now make use of this fact.

To generate a class of NCQ mechanics replace each $\delta(x)$ in (\ref{equation 1}) with \emph{any} positive semidefinite delta sequence $\delta_m (x)$, as an example one may take the Gaussian delta sequence, $\delta_m(x)=\frac{m}{\sqrt \pi}e^{-m^2x^2}$. The necessity of the positive semidefinite condition arises in order to keep the probability interpretation. This procedure yields a class of PDFs, \begin{eqnarray}\label{II}P_{\delta_m}[\mathbf{x}(t)]=F[\mathbf{x}(t)] \int d\alpha_1 d\alpha_2...d\alpha_n \label{I} \prod_t \delta_m(\mathbf{x}(t)-\mathbf{x}_s(\alpha_1, \alpha_2,...\alpha_n;t))\end{eqnarray} where each element of the class corresponds to a particular delta sequence and a particular value for the delta parameter $m$. Since $\lim_{m\to \infty} \delta_m(x)=\delta(x)$, $\lim_{m\to \infty} P_{\delta_m}[\mathbf{x}(t)]=P_c[\mathbf{x}(t)]$, which is to say that each $P_{\delta_m}[\mathbf{x}(t)]$ has a well defined classical limit. Generally $\delta_m(x)\not=0$, as it is with our Gaussian example. Consequently, $P_{\delta_m}[\mathbf{x}(t)]$ is generally nonzero \emph{for} \emph{all} $\mathbf{x}(t)$, and so all trajectories are hypothetically observable. We see that classical mechanics is replaced by a stochastic process.

The replacement of the delta functions with delta sequences is hitherto arbitrary. Within the context of this note, the choice of delta sequence and the value of the delta parameter must be chosen so as to agree with experiment. There is, however, another consideration that suggests a particular delta sequence. An important feature of a quantum mechanical wavefunction is the existence of nodes and anti-nodes, or points of ``minimum and maximum probability,'' respectively, and of course the generation of interference effects. In order for NCQ mechanics to yield nodal or (non-classical) anti-nodal paths, it is necessary to choose a delta sequence that exhibits these features. Thus, one should consider the following delta sequence $\frac{1}{\pi}\frac{\sin^2mx}{mx^2}$ \cite{refSakurai} which satisfies \begin{equation}\lim_{m\to \infty}\frac{1}{\pi}\frac{\sin^2mx}{mx^2}=\delta(x),\end{equation} is positive semidefinite, and exhibits (anti-)nodes away from the classical path. To the author's knowledge this is the only delta sequence that meets these criteria. Furthermore, it is the belief of the author that symmetry considerations in a theory more fundamental than that of NCQ mechanics will eliminate the need for choosing a delta sequence. 

These considerations suggest that the appropriate form of the PDF is \begin{equation}\label{III}P_{m}[\mathbf{x}(t)]=F[\mathbf{x}(t)] \int d\alpha_1...d\alpha_n \label{I}\prod^3_{i=1}  \prod_t\frac{1}{\pi}\frac{\sin^2m[(x^i(t)-x^i_s(\alpha_1,...\alpha_n;t)]}{m[(x^i(t)-x^i_s(\alpha_1,...\alpha_n;t)]^2}\phantom{iiii}\end{equation} where $m$ is fixed by comparison with experiment. Using (\ref{III}) every classical mechanical system with a known general solution has a corresponding NCQ mechanical description. However, the structure of (\ref{III}) makes explicit computations formidable. It is for this reason that the calculation of quantities that may be directly compared with experiment is beyond the scope of this introductory note.   

NCQ mechanics has notable features in common with both classical and quantum mechanics. In the aforementioned class of NCQ mechanics, each particle has a position at each time, and follows a well defined path. In this way, NCQ mechanics exhibits a key characteristic of CM. It exhibits several features unique to QM. Clearly it introduces probability in an inherent way in the theory from the onset, like QM. Fixing the initial conditions still corresponds to making a choice for the normalizing function $F[\mathbf{x}(t)]$, however, since all trajectories are observable, fixing initial conditions is \emph{not} sufficient to determine what trajectory a particle will follow. NCQ mechanics is indeterministic. To see that it maintains an even deeper analogy with QM consider the following. If $P[\mathbf{x}(t)]$ is taken to equal the classical PDF, $P_c[\mathbf{x}(t)]$, then in the expectation value of any arbitrary observable $O$ only the classical trajectories contribute, as expected. However, in the case of NCQ mechanics, with $P=P_{\delta_m}[\mathbf{x}(t)]$, it is generally true that all trajectories contribute to $\langle O \rangle$. That this is part and parcel of QM is most obviously seen in the path integral formalism where the expectation value of an observable has nonzero contributions coming from all trajectories  \cite{refFeynman}. Making use of the classical limit in NCQ mechanics (large delta parameter) and the classical limit of QM we have the following statements: \begin{equation}\lim_{m \to \infty} \langle O \rangle_{NCQ}=\langle O \rangle_c\end{equation} \begin{equation}\lim_{\hbar \to 0} \langle O \rangle_{QM}=\langle O \rangle_c\end{equation}   

\section{Summary and Conclusions}
\label{sec:4}

In this note we have argued for the benefits of studying general NCQs, or theories that differ from both classical and quantum theories, but which exhibit characteristics of both, because of the possibility that if appropriately chosen they may represent alternatives to standard quantum theory. To this end we introduced a reformulation of classical mechanics in terms of a probability density functional, $P_c[\mathbf{x}(t)]$. It assigned a nonzero probability (read probability density) to all classically allowed trajectories, and is completely fixed given the initial conditions and the equations of motion. From this classical PDF we defined the classical expectation value of an observable. We observed that $P_c[\mathbf{x}(t)]$ contained an integrated product of delta functions, see (\ref{equation 2}). The delta functions exactly restrict the PDF to be nonzero on the classically allowed trajectories.

NCQ mechanics was arrived at by replacing each delta function with a positive semidefinite delta sequence. For each such delta sequence and for each value of the delta parameter $m$, one finds a new element in a class of NCQ mechanics theories. Since each delta sequence may be made arbitrarily close to a delta function with the adjustment of the delta parameter, NCQ mechanics is necessarily endowed with a classical limit. NCQ mechanics shares key characteristics with both QM and CM. Particles have a position at all times and follow trajectories in NCQ mechanics. Yet, probabilities appear inherently from the onset in the theory, it is indeterministic, and all trajectories contribute to the expectation values of observables.

\end{document}